\documentclass[twocolumn,showpacs,superscriptaddress,prl]{revtex4}

\usepackage{graphicx}
\usepackage{amsmath}
\usepackage{amsfonts}
\usepackage{color}
\usepackage{bm}
\usepackage[normalem]{ulem}

\newcommand{\ket}[1]{\ensuremath{\left|{#1}\right\rangle}}
\newcommand{\bra}[1]{\ensuremath{\left\langle{#1}\right |}}

\newcommand{\beq}{\begin{equation}}
\newcommand{\eeq}{  \end{equation}}
\newcommand{\bea}{\begin{eqnarray}}
\newcommand{\eea}{  \end{eqnarray}}
\newcommand{\bit}{\begin{itemize}}
\newcommand{\eit}{  \end{itemize}}

\graphicspath{{converted_graphics/}}

\newcommand{\ie}{{\it{i.e.}}}

\begin{document}

\title{Linear-optical simulation of the cooling of a cluster-state Hamiltonian system}
\author{G. H. Aguilar}
\affiliation{Instituto de F\'{\i}sica, Universidade Federal do Rio de Janeiro, Caixa
Postal 68528, Rio de Janeiro, RJ 21941-972, Brazil}
\author{T. Kolb}
\affiliation{Instituto de F\'{\i}sica, Universidade Federal do Rio de Janeiro, Caixa
Postal 68528, Rio de Janeiro, RJ 21941-972, Brazil}
\author{D. Cavalcanti}
\affiliation{ICFO-Institut de Ciencies Fotoniques, 08860 Castelldefels (Barcelona), Spain}
\author{L. Aolita}
\affiliation{Dahlem Center for Complex Quantum Systems, Freie Universit\"{a}t Berlin, Berlin, Germany}
\author{R. Chaves}
\affiliation{Institute for Physics, University of Freiburg, Rheinstrasse 10, D-79104 Freiburg, Germany}
\author{S. P. Walborn}
\affiliation{Instituto de F\'{\i}sica, Universidade Federal do Rio de Janeiro, Caixa
Postal 68528, Rio de Janeiro, RJ 21941-972, Brazil}
\author{P. H. Souto Ribeiro}
\affiliation{Instituto de F\'{\i}sica, Universidade Federal do Rio de Janeiro, Caixa
Postal 68528, Rio de Janeiro, RJ 21941-972, Brazil}
\date{\today }
\begin{abstract}
A measurement-based quantum computer could consist of a local-gapped Hamiltonian system, whose thermal states --at sufficiently low temperature-- are universal resources for the computation. Initialization of the computer would correspond to cooling the system. We perform an experimental quantum simulation of such cooling process with entangled photons.
We prepare three-qubit thermal cluster states exploiting the equivalence between local dephasing and thermalisation for these states. This allows us to tune the system's temperature by changing the dephasing strength.
We monitor the entanglement as the system cools down and observe the transitions from separability to bound
entanglement, and then to free entanglement. We also analyze the performance of the system for measurement-based single-qubit state preparation. These studies constitute a basic characterisation of experimental cluster-state computation under imperfect conditions.
\end{abstract}

\maketitle




\par \textit{Introduction:} One of the main approaches to quantum computing is the measurement-based quantum computation (MBQC) model \cite{oneway01}. There, computations are performed by adaptive single-particle measurements on lattice systems prepared in many-body entangled states, which are universal resources for the computation. Particularly convenient is the case when such states happen to be the unique ground state of a gapped 
Hamiltonian. Then, the resource for the computation is obtained by cooling the Hamiltonian system down to its ground state. The energy gap $\Delta$, in turn, provides an intrinsic energy barrier against thermal excitations that may drive the system out of the ground state. Several examples of these Hamiltonians have been found for interacting spin \cite{Chen09, Cai10, Wei11} and bosonic \cite{Aolita11, Menicucci11} systems.

The best studied example, and the only one known for the case of qubits \cite{Bartlett06, Chen11}, is the cluster-state Hamiltonian. Its ground state, as the name suggests, is the cluster state, which is a universal resource for the one-way model of MBQC \cite{oneway01}. It belongs to the more general family of graph states, which possess a variety of applications in quantum information and communication \cite{Hein04,Hein06}. In addition, fault-tolerant error correction is fully developed for the cluster-state one-way model \cite{Raussendorf06,Raussendorf07}. In particular, not only the ground state but also the thermal states up to a temperature of the order of $\Delta$ are universal resources for MBQC \cite{Raussendorf05}. Furthermore, this Hamiltonian can in principle be efficiently cooled down by local interaction with independent thermal baths at low temperature \cite{Jennings09, Aolita13}. However, such cooling process is still pending experimental demonstration. Thus, quantum simulations constitute a powerful tool for the experimental study of this kind of system \cite{orieux}.

In this work, we experimentally study three-qubit thermal cluster states at tunable temperatures. We use the polarization of two photons to encode two of the qubits and a path degree of freedom of one of the photons to encode the third one. We exploit the equivalence between local dephasing and thermalisation for cluster states \cite{Raussendorf05, kay06, dcavalcanti10}, which allows us to tune the system's temperature by changing the dephasing strength.
 We perform state tomography for a sequence of temperatures ranging from high temperatures, corresponding to complete dephasing, to nearly zero, corresponding to almost pure entangled states. The fidelities of the experimental states with respect to the ideal thermal states are above $93\%$. 
  Implementing a real cluster-state Hamiltonian is still challenging. We produce 
the corresponding thermal states with a quantum simulation (see Ref. \cite{walther12} and the references therein). We consider the case where the simulated system is in contact with a reservoir whose temperature decreases adiabaticaly, so that system and reservoir remain in thermal equilibrium throughout the temperature change. We monitor the entanglement as the temperature decreases. Within the experimental uncertainties, it is possible to observe the  transition from separability to bound entanglement, and subsequently from bound  to distillable entanglement, as is predicted theoretically \cite{dcavalcanti10, kay10}. Interestingly, the medium-temperature states created are, to our knowledge, the first experimental observation of both thermal bound entangled states and of bound entangled states of three qubits. Finally, to analyse the effects of temperature on the thermal linear-cluster state as a computational resource, we implement a measurement-based state preparation and measure its average fidelity over generic single-qubit target states.

\par \textit{Theory:} Cluster states correspond to graph states whose associated graph is a rectangular lattice. A general $N$-qubit graph state $\ket{G^{\boldsymbol{0}}_{N}}$ is associated to a graph $G$, composed of $N$ vertices and a set $E$ of edges $\{i,j\}$ connecting vertices $i$ and $j$ for $1\leq i, j\leq N$, which determines the geometry of $G$. The usual operational definition is  \cite{Hein04,Hein06}
\beq
\label{ground_state}
\ket{G^{\boldsymbol{0}}_{N}}\doteq\prod_{\{i,j\}\in E} CZ_{ij} \ket{+}^{\otimes N},
\eeq
where $\ket{+}=(\ket{0}+\ket{1})/\sqrt{2}$, and $\ket{0}$ and $\ket{1}$ are the computational-basis states. The operation $CZ_{ij}\doteq(\ket{0_i}\bra{0_i}\otimes \openone_j + \ket{1_i}\bra{1_i}\otimes Z_j)\otimes\openone_{\overline{ij}}$, with $Z_j$ the third Pauli operator acting on qubit $j$, and $\openone_j$ and $\openone_{\overline{ij}}$ the identity operators on qubits $j$ and all qubits but $i$ and $j$, respectively, is the maximally entangling controlled-Z gate acting non-trivially on qubits $i$ and $j$.

\par An alternative definition of  \eqref{ground_state} is through its {\it parent Hamiltonian}
\beq
H\doteq-\frac{\Delta}{2}\sum_{i=1}^{N} X_i \bigotimes_{j\in\mathcal{N}_i} Z_j,
\label{Hamiltonian}
\eeq
with $\Delta > 0 $ the {\it energy gap}, $X_i$ the usual first Pauli operator acting on qubit $i$, and $\mathcal{N}_i$ the set of first-neighbours of qubit $i$ according to $E$. All $N$ local operators appearing in summation \eqref{Hamiltonian} commute and have eigenvalues 1 or $-1$. The Hamiltonian has then a unique ground state of eigenenergy $-\frac{N \Delta}{2}$ and is in addition {\it frustration-free}, meaning that the ground state of the total Hamiltonian is also the ground state of each local term in the sum. The unique ground state is nothing but $\ket{G^{\boldsymbol{0}}_{N}}$.

The eigenstates of \eqref{Hamiltonian} can be written as $\ket{G_{N}^{\boldsymbol{\mu}}}\doteq\bigotimes_{i=1}^N Z_i^{\mu_i} \ket{G^{\boldsymbol{0}}_{N}}$, where $\boldsymbol{\mu}\doteq( \mu_1,\ \dots \mu_N)$, with $\mu_i =0$ or $1$, for all $1\leq i\leq N$, and have eigenenergies $-\frac{\Delta}{2}\sum_{i=1}^N (-1)^{\mu_i}$. That is, $Z_i$ creates an excitation of $H$ on $\ket{G^{\boldsymbol{0}}_{N}}$. This can be seen from the fact that $Z_i X_i \bigotimes_{j\in\mathcal{N}_i} Z_j Z_i=-X_i \bigotimes_{j\in\mathcal{N}_i} Z_j$, for all $i$ and that $H$ is a commuting frustration-free Hamiltonian.
The multi-index $\boldsymbol{\mu}$ can be thought of as an excitation vector, whose norm gives the number of excitations. The energy difference between the ground state and the first-excited manifold is $\Delta$, which explains the name ``gap". Accordingly, the thermal state $\rho_T$ at equilibrium temperature $T$ (in units of  Boltzman's constant $k_B$) is defined as
\beq
\rho_T=\frac{e^{-H/T}}{\mathrm{Tr}[{e^{-H/T}]}}.
\label{thermal_state}
\eeq

Since system excitations are created by the $Z$ Pauli operators, the thermal state is equivalent to the ground state under independent dephasing  \cite{Raussendorf05, kay06, dcavalcanti10}. That is,
\beq
\rho_T\equiv\mathcal{E}_1 \otimes \mathcal{E}_2 \dots \otimes \mathcal{E}_N \ket{G^{\boldsymbol{0}}_{N}}\bra{G^{\boldsymbol{0}}_{N}},
\label{thermal_state2}
\eeq
where
\beq
\mathcal{E}_i\rho=\left(1-\frac{p}{2}\right)\rho+\frac{p}{2} Z_i \rho Z_i
\label{canal_deph}
\eeq
is the dephasing channel on qubit $i$, for any state $\rho$, with dephasing strength
\beq
p=\frac{2}{1+e^{\Delta/T}}.
\label{TvsP}
\eeq


\textit{Experiment:} In our experiment, we use equivalence \eqref{thermal_state2} to create $\rho_T$ for a 1D graph of $N=3$, \ie~the 3-qubit thermal linear-cluster state. More precisely, we experimentally prepare entangled photons in (an almost pure) three-qubit linear-cluster state and apply independent dephasing on each qubit. As we can see from Eq. \eqref{TvsP}, the temperature of the thermal state is tuned by choosing the strength $p$ of the dephasing channel.

\begin{figure}
\centering
\includegraphics[width=8.5cm]{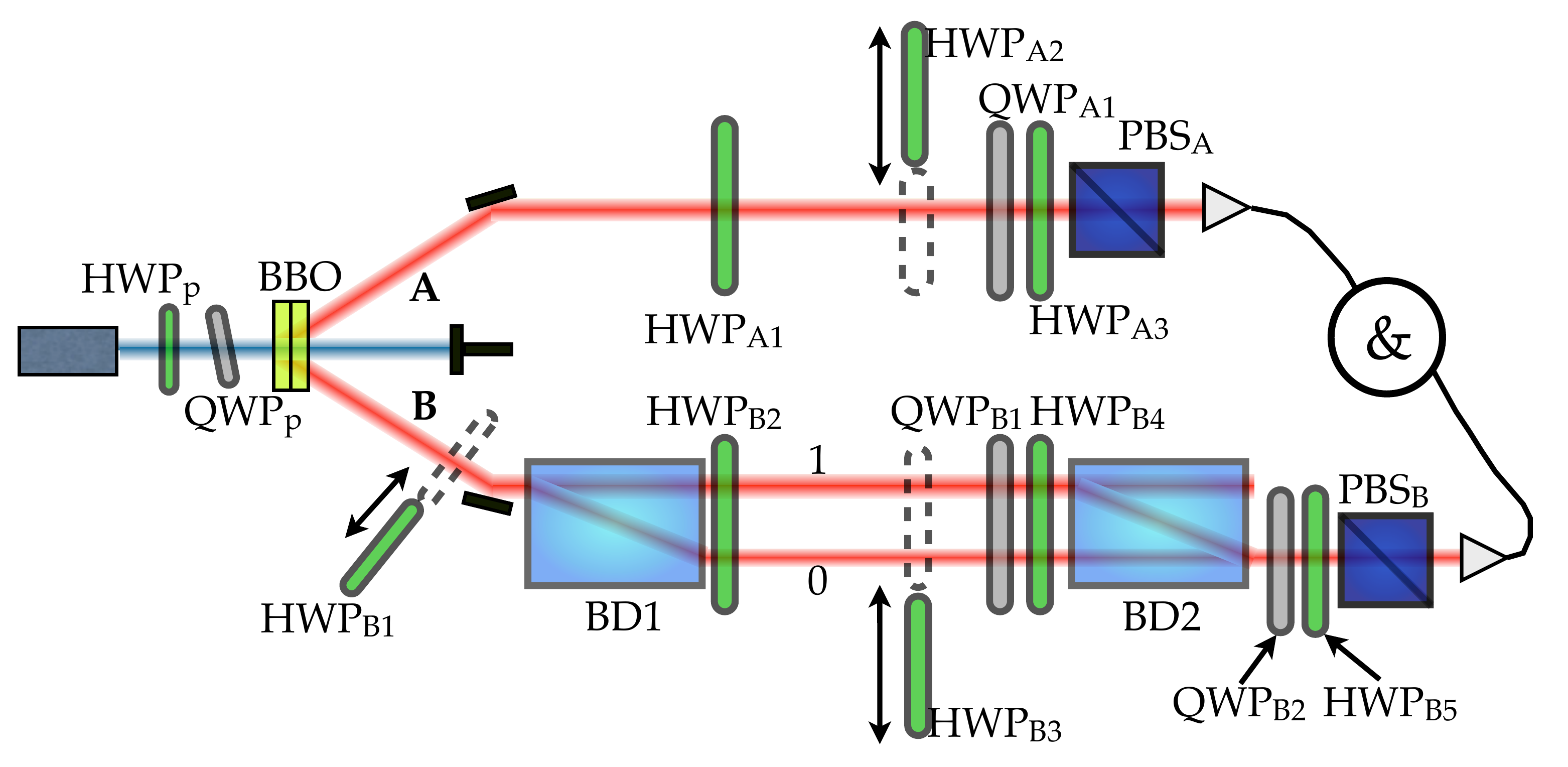}
\caption{Experimental setup. Experimental setup. HWP is half wave plate, QWP is quarter wave plate, BBO is Barium beta borate
non-linear crystal, BD is beam displacer, and PBS is
polarizing beam-splitter. See text for more detail. 
}
\label{fig:configuration}
\end{figure}

The experimental setup is shown in Fig. \ref{fig:configuration}. A He-Cd (Helium Cadmium) laser at 325 nm pumps two cross-axis BBO crystals and produces entangled photons at $650$ nm. The state of these photons can be written as \cite{kwiat99}:
\begin{equation}
\label{polentangle}
\ket{\Phi}=\frac{1}{\sqrt{2}}\left( \ket{0}_{\text{A}_\text{p}}\ket{0}_{\text{B}_\text{p}}+ \ket{1}_{\text{A}_\text{p}}\ket{1}_{\text{B}_\text{p}}\right),
\end{equation}
where $\ket{0}$ ($\ket{1}$) is the horizontal (vertical) polarization of each photon; and the labels $\text{A}_\text{p}$ and $\text{B}_\text{p}$ refer to the qubits encoded in the polarization of photons A and B, respectively.
The photons produced in the spatial mode A are detected after polarization analysis. Photon B is sent to a beam displacer (BD1) that, due to its birefringence, transmits the $V$-polarized photons in spatial mode $1$  
and deflects the $H$-polarized photons to mode $0$. Due to BD1, the spatial degree of freedom of photon B becomes entangled with the photon's polarization, producing the state
$\ket{\Psi}=(\ket{0}_{\text{A}_\text{p}}\ket{0}_{\text{B}_\text{s}}\ket{0}_{\text{B}_\text{p}}+ \ket{1}_{\text{A}_\text{p}}\ket{1}_{\text{B}_\text{s}}\ket{1}_{\text{B}_\text{p}})/\sqrt{2}$, where $\text{B}_\text{s}$ labels the spatial-mode qubit of photon B.
To obtain state  \eqref{ground_state}, we apply a Hadamard gate, which maps $\ket{0}$ into $\ket{+}$ and $\ket{1}$ into $\ket{-}$, on qubits $\text{A}_\text{p}$ and $\text{B}_\text{p}$.
This is done with half-wave plates HWP$_{\text{A}1}$ and HWP$_{\text{B}2}$, which finally lead to the desired state:
\beq
\ket{G^{\boldsymbol{0}}_{3}}=\frac{1}{\sqrt{2}}(\ket{+}_{\text{A}_\text{p}}\ket{0}_{\text{B}_\text{s}}\ket{+}_{\text{B}_\text{p}}+ \ket{-}_{\text{A}_\text{p}}\ket{1}_{\text{B}_\text{s}}\ket{-}_{\text{B}_\text{p}}).
\label{ground_state_2}
\eeq

Fig. \ref{fig:configuration} also describes the detection setup used for the tomographic reconstruction of the experimental three-qubit state. Since photons  A encode only polarization qubits, tomographic measurements are performed as usual, using a quarter-waveplate QWP$_{\text{A}1}$, a half-waveplate HWP$_{\text{A}3}$, and a polarizing beam-splitter PBS$_{\text{A}}$ \cite{james01}. For photons  B, tomographic measurements require measuring the polarization and spatial-mode degrees of freedom simultaneously. Here we make use of the same configuration as in Ref. \cite{farias12b}. With the waveplates QWP$_{\text{B}1}$ and HWP$_{\text{B}4}$, and the beam displacer BD2, we perform tomographic measurements on qubit B$_p$.
As we can see in Fig. \ref{fig:configuration}, BD1 and BD2 form an interferometer, so that paths 0 and 1 are recombined coherently at BD2. This is crucial for the tomographic measurements of qubit B$_s$.
Since BD2 also deflects photons with polarization $0$ and transmits those with polarization $1$, the $0$-polarized photons at the interferometer output correspond to those of path $0$ inside the interferometer. In the same way, photons in path $1$ inside the interferometer are detected at the output with polarization  $1$. Thus, the tomographic measurements of B$_s$ are performed with QWP$_{\text{B}2}$, HWP$_{\text{B}5}$ and PBS$_{\text{B}}$, outside the interferometer. Then, the photons are coupled into single-mode fibers that are connected to single-photon detectors, and coincidence events are registered. To perform the reconstruction of the complete three-qubit density matrix, 64 settings of QWP$_{\text{A}1}$, HWP$_{\text{A}3}$, QWP$_{\text{B}1}$, HWP$_{\text{B}4}$, QWP$_{\text{B}2}$,  and HWP$_{\text{B}5}$ are used, following the standard recipe \cite{james01}.

Thermal states \eqref{thermal_state} are created by applying the dephasing channel \eqref{canal_deph} to each qubit of \eqref{ground_state_2}, according to equivalence \eqref{thermal_state2}. Channel \eqref{canal_deph} is a sum of two events: i) with probability $(1-p/2)$, the state remains unchanged, and ii) with probability $p/2$, a $Z$ gate is applied to the qubit. This is implemented by toggling in and out the three half-waveplates HWP$_{\text{A}2}$, HWP$_{\text{B}1}$, and HWP$_{\text{B}3}$, which act as $Z$ gates on qubits A$_p$, B$_s$ and B$_p$, respectively. Notice that the $Z$ gate on qubit B$_s$ is implemented by HWP$_{\text{B}1}$, inserted before the interferometer. This is due to the fact that a relative $\pi$ phase between polarizations 0 and 1 before the interferometer is equivalent to a relative $\pi$ phase between the paths  0 and 1 inside the interferometer. The three half-waveplates are inserted in the path of both photons a fraction $p/2$ of the total measurement runs.
In this way, by averaging over the outcomes of all measurement runs, one effectively implements the desired dephasing channel \eqref{canal_deph} on all three qubits.

\begin{figure}
\centering
\includegraphics[width=6.5cm]{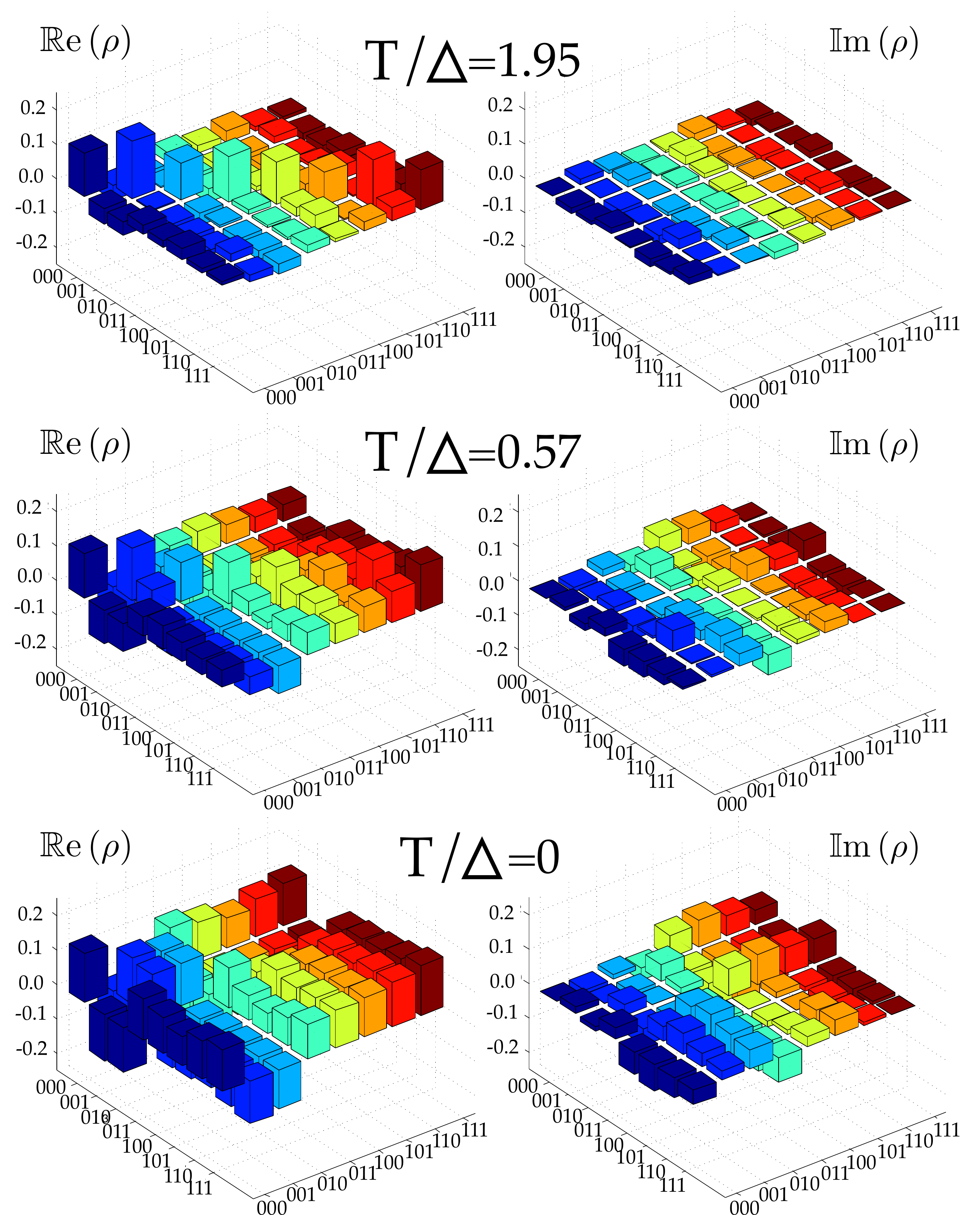}
\caption{ Real (left panels) and imaginary (right panels) parts of the reconstructed density matrices for three different temperatures. The ground state $\ket{G^{\boldsymbol{0}}_{3}}$ corresponds to $T/\Delta=0$. As  $T/\Delta$ increases, the coherences decrease, until $T/\Delta=1.95$, when they become negligible and the state is practically the maximally mixed.}
\label{fig:matrices}
\end{figure}

\textit{Results:} We tomographically reconstructed the three-qubit density matrices for many values of $p$, or equivalently $T$.  Fig. \ref{fig:matrices} shows the density matrices reconstructed for three values of $T/\Delta =1/\left( \ln(2-p)-\ln p\right)$. The produced states have fidelities larger than $0.93\%$ with thermal states \eqref{thermal_state} at temperature given by \eqref{TvsP}.

\begin{figure}
\centering
\includegraphics[width=7.3cm]{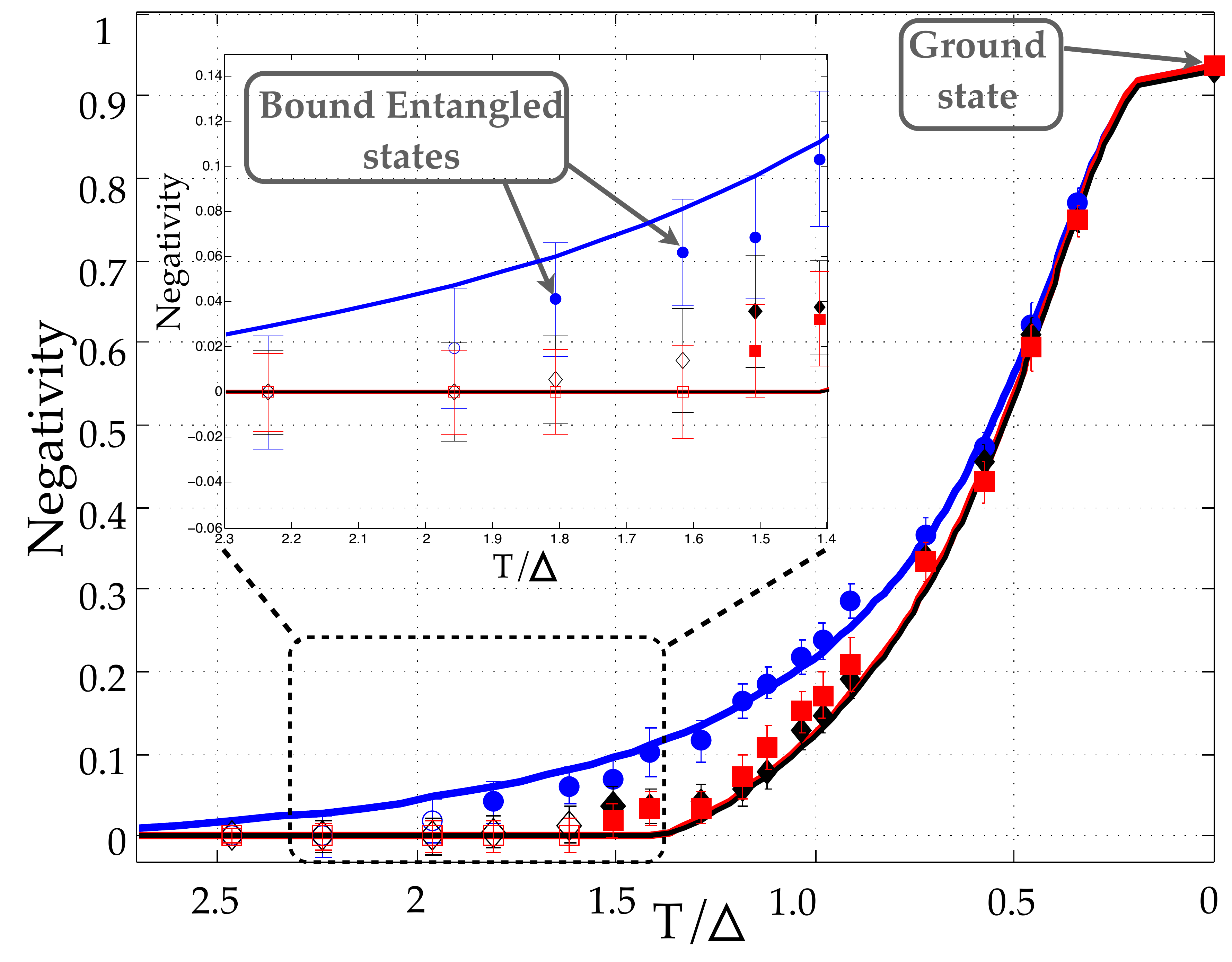}
\caption{Negativities $N_{\text{A}_\text{p}|\text{B}_\text{s}\text{B}_\text{p}}$ (red squares), $N_{\text{B}_\text{p}|\text{A}_\text{p} \text{B}_\text{s}}$ (black diamonds) and  $N_{\text{B}_\text{s}|\text{A}_\text{p} \text{B}_\text{p}}$ (blue circles), for the three bipartitions $\text{A}_\text{p}|\text{B}_\text{s}\text{B}_\text{p}$, $\text{B}_\text{p}|\text{A}_\text{p} \text{B}_\text{s}$ and  $\text{B}_\text{s}|\text{A}_\text{p} \text{B}_\text{p}$, respectively, of the experimental density matrices, as functions of $T/\Delta$. The solid lines correspond to the negativities of the theoretical thermal cluster states, with the substitution of $Z$ with $F(\alpha)$ (see text). As the temperature increases, $N_{\text{A}_\text{p}|\text{B}_\text{s} Bp}$ and $N_{\text{B}_\text{p}|\text{A}_\text{p} \text{B}_\text{s}}$ vanish before $N_{\text{B}_\text{s}|\text{A}_\text{p} \text{B}_\text{p}}$, as can be seen in the inset. This causes the emergence of bound entangled states (see text). We symbolize null negativities  with hollow symbols.}
\label{fig:results}
\end{figure}

Due to experimental imperfections, the $Z$ gates in the implementation of the dephasing channels are not ideal. A better description of the reconstructed density matrices corresponds to a modified dephasing channel, where the $Z$-gate is replaced by the phase-gate $F(\alpha)=\ket{0}\bra{0}+e^{i\alpha}\ket{1}\bra{1}$. A good agreement with the experimental data is obtained with $\alpha=0.84\pi$, as can be seen in Fig. \ref{fig:results}.

Let us now analyze the entanglement dependence on the temperature. In Fig \ref{fig:results} we can see the evolution of the bipartite entanglement, as measured by the negativity \cite{vidal02}, across the three bipartitions, as we cool down the system. As we can see, for high temperatures $T/\Delta\gtrsim1.95$, the negativities in all three partitions are zero. At $T/\Delta\simeq1.95$, negativity $N_{\text{B}_\text{s}|\text{A}_\text{p} \text{B}_\text{p}}$, with respect to qubit $\text{B}_\text{s}$, becomes positive, while $N_{\text{A}_\text{p}|\text{B}_\text{s} \text{B}_\text{p}}$ and $N_{\text{B}_\text{p}|\text{A}_\text{p} \text{B}_\text{s}}$, with respect to qubits $\text{A}_\text{p}$ and $\text{B}_\text{p}$, respectively, remain null, up to the experimental uncertainty.
This difference comes from the fact that state \eqref{ground_state_2} is not symmetric with respect to the exchange of qubits. 
At this point, the system becomes  bound entangled \cite{dcavalcanti10}. If $N_{\text{A}_\text{p}|\text{B}_\text{s} \text{B}_\text{p}}=0=N_{\text{B}_\text{p}|\text{A}_\text{p} \text{B}_\text{s}}$, no entanglement can be extracted from the bipartitions  ${\text{A}_\text{p}|\text{B}_\text{s} \text{B}_\text{p}}$ or ${\text{B}_\text{p}|\text{A}_\text{p} \text{B}_\text{s}}$ by local operations assisted by classical communication (LOCCs). Consequently, no entanglement can be distilled between any two qubits by individual local operations at each qubit, because any pair of qubits is splitted either by biparition $\text{A}_\text{p}|\text{B}_\text{s} \text{B}_\text{p}$ or by $\text{B}_\text{p}|\text{A}_\text{p} \text{B}_\text{s}$, both of which are PPT \cite{dcavalcanti10}. This implies that no entanglement can be extracted at all by individual LOCCs. However, the negativity for the partition $\text{B}_\text{s}|\text{A}_\text{p} \text{B}_\text{p}$ is  positive, meaning that the system is entangled. This kind of multipartite bound entanglement was observed for four qubits in Refs. \cite{amselem09, lavoie10,barreiro10,Kaneda2012}.
The emergence of bound entanglement is better appreciated in the inset. There, one observes for instance that, at temperatures $T/\Delta\simeq1.8$ and $T/\Delta\simeq1.61$,  $N_{\text{A}_\text{p}|\text{B}_\text{s} \text{B}_\text{p}}$ and $N_{\text{B}_\text{p}|\text{A}_\text{p} \text{B}_\text{s}}$ are null within the error bars (of size 0.02), whereas $N_{\text{B}_\text{s}|\text{A}_\text{p} \text{B}_\text{s}}$ is respectively $0.04 (0.02)$ and $0.06 (0.02)$ for these temperatures. Finally, in the low-temperature region $T/\Delta\lesssim1.5$, all the negativities are positive and the entanglement is thus distillable.

In the procedure for the calculation of the error bars we assume a Poissonian distribution
for the coincidence counts and perform Monte Carlo simulation to obtain a distribution of
negativities, and take its standard deviation as the error. The small error bars were achieved thanks to a relatively high coincidence counting rates (200 per second at the populations) and large sampling time for every projective measurement.  The measured bound entangled state at temperature $T/\Delta\simeq1.8$ has a fidelity $\mathcal{F} = 0.95$ with respect to the theoretically predicted state. To our knowledge, this is the first time a bound entangled state is measured in a three-qubit system.

Finally, we study the usefulness of our thermal states as resources for noisy MBQC \cite{Chaves_noisy_2011}. In particular, we implement a measurement-based single-qubit state preparation, and measure its average fidelity over Haar-random single-qubit target states. The latter is equivalent to the average over any two-design (see \cite{Renes04} and Refs. therein, for example). This is very convenient, as two-designs are for instance given by the eigenstates in any set of mutually unbiased bases \cite{Klappenecker05}. We choose the mutually unbiased bases given by the eigenstates of the $X$, $Y$ and $Z$ Pauli operators:  $\{\ket{+}, \ket{-}\}$, $\{\ket{r}\doteq(\ket{0}+ i\ket{1})/\sqrt{2},\ket{l}\doteq(\ket{0}-i\ket{1})/\sqrt{2}\}$ and $\{\ket{0}, \ket{1}\}$, respectively.

\begin{figure}
\centering
\includegraphics[width=7.3cm]{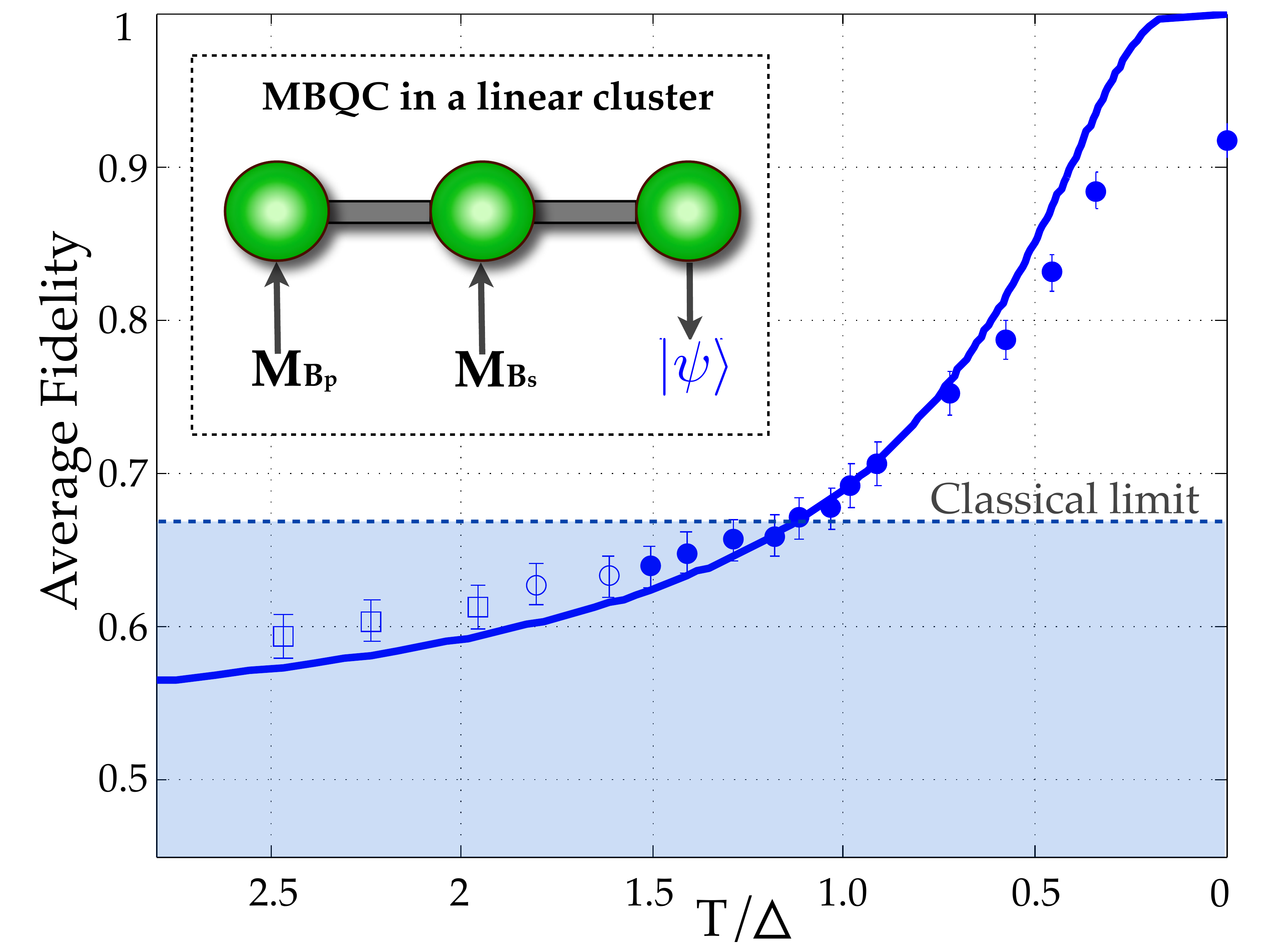}
\caption{Experimental results:  Average fidelity in the single-qubit state preparation (see inset) as a function of the temperature. Hollow squares indicate non-distillable (probably separable) states, and hollow circles bound entangled states (non-distillable with respect to just two bipartitions). Solid circles, in turn, correspond to free entangled states.}
\label{fig:fidelities}
\end{figure}

The protocol is schematically shown in the inset of Fig. \ref{fig:fidelities}. We make projective measurements $M_{\text{B}_\text{p}}$ and $M_{\text{B}_\text{s}}$ in either the $X$, $Y$ or $Z$ bases, preparing in each case a conditional state $\varrho_{\text{A}_\text{p}}$ in qubit $\text{A}_\text{p}$. Ideally, $\varrho_{\text{A}_\text{p}}$ should be an eigenstate of $X$, $Y$ or $Z$. The desired average fidelity is thus obtained by averaging the fidelity of $\varrho_{\text{A}_\text{p}}$ with the expected eigenstate for each measurement choice and outcome, according to usual cluster-state computation \cite{oneway01}. The protocol is repeated for different temperatures.

The experimental results are shown in Fig.  \ref{fig:fidelities}. The solid line is a theoretical curve obtained by taking the ideal initial state with unity purity and
evolving it to higher temperatures through the dephasing channel, just like in Eq. \eqref{thermal_state2}. The average fidelity of the prepared states surpasses the
classical benchmark of $2/3$ \cite{Horodecki1999} for temperatures $T/\Delta \lesssim 1.1$, showing the usefulness of these thermal states. We observe a good agreement between theory and experiment
for intermediate temperatures between $0.5 \gtrsim T/\Delta \gtrsim 2$. For $T/\Delta \lesssim0.5$, the experimental points are shifted from the theoretical curve. Even though the initial state is highly pure, the small amount
of mixedness might be responsible for this deviation. Below the classical limit, the experimental fidelities are higher than the theoretical predictions, possibly due to residual classical correlations between the polarization and path degrees of freedom in the same photon.

\textit{Conclusions:}
Low-temperature states of systems governed by experimentally feasible Hamiltonians offer a promising platform for universal measurement-based quantum computation. Experimental investigations of the role of thermalization, with emphasis on the nature of entanglement decay it induces on these systems and the limitations it imposes on them as computational resources, are very timely. Here, we performed such an investigation with entangled photons and linear-optical networks. We prepared three-qubit thermal linear-cluster states at tunable temperature. This allowed us to simulate the system cooling process.

We characterized the entanglement dynamics as temperature decreases, observing the transition from separability, at infinite temperature, to bound entanglement, at intermediate temperatures, and finally to free, distillable  entanglement, for low temperatures. Experimental bound entanglement has already been reported \cite{amselem09, lavoie10, barreiro10, DiGuglielmo2011, Kaneda2012, Hiesmayr2013}. However, the bound entangled states reported here constitute, to our knowledge, the first experimental observation of this type of entanglement both in thermal states and in the lowest-dimensional system for which this kind of entanglement is possible.

Finally, we analyzed the effects of non-zero temperature in a simple, exemplary cluster-state computation: the preparation of an arbitrary single-qubit state. We characterized the range of temperatures for which the thermal states provide average fidelities higher than those attainable with any classical strategy.

These studies give a basic experimental characterization of the dynamics of cluster-state systems in thermal equilibrium with a bath whose temperature varies slowly. Our proof-of-principle photonic implementation provides thus useful grounds for future experimental studies of similar systems with more general physical platforms.

\begin{acknowledgements}
Financial support was provided by Brazilian agencies CNPq, CAPES
FAPERJ, and the Instituto Nacional de Ci\^encia e Tecnologia - Informa\c{c}\~ao Qu\^antica. LA acknowledges the support from the EU under Marie Curie IEF No 299141, RC from the Excellence Initiative of the German Federal and State Governments (Grant ZUK 43), and DC from the EU project SIQS.
\end{acknowledgements}



\begin{thebibliography}{34}

\bibitem{oneway01}
  R. Raussendorf and H. J. Briegel,
  Phys. Rev. Lett.
 \textbf{86},
 5188
  (2001).
  
  \bibitem{Chen09}
	X. Chen, B. Zeng, Z. C. Gu, B. Yoshida and I. L. Chuang,
	Phys. Rev. Lett.
	\textbf{102},
	220501
	(2009).

\bibitem{Cai10}
	J. Cai, A. Miyake, W. D\"ur and H. J. Briegel,
	Phys. Rev. A
	\textbf{82},
	052309
	(2010).
	
\bibitem{Wei11}
	T.-C. Wei,
	Phys. Rev. Lett.
	\textbf{106},
	070501
	(2011).


\bibitem{Aolita11}
	L. Aolita, A. J. Roncaglia, A. Ferraro and A. Ac\'in,
	Phys. Rev. Lett.
	\textbf{106},
	090501
	(2011).


\bibitem{Menicucci11}
	N. C. Menicucci, S. T. Flammia and P. van Loock,
	Phys. Rev. A
	\textbf{83},
	042335
	(2011).
	
	


\bibitem{Bartlett06}
	S. D. Bartlett and T. Rudolph,
	Phys. Rev. A
	\textbf{74},
	040302(R)
	(2006).
	
	
	
\bibitem{Chen11}
	J. Chen, X. Chen, R. Duan, Z. Ji and B. Zeng,
	Phys. Rev. A
	\textbf{83},
	050301
	(2011).
	
	
\bibitem{Hein04}
	M. Hein, J. Eisert and H. J. Briegel,
	Phys. Rev. A
	\textbf{69},
	062311
	(2004).
	
\bibitem{Hein06}
	M. Hein, W. D\"{u}r, J. Eisert, R. Raussendorf, M. Van den Nest and H. J. Briegel,
	arXiv: quant-ph/0602096
	(2006).

\bibitem{Raussendorf06}
	R. Raussendorf, J. Harrington and K. Goyal,
	Annals of Phys.
	\textbf{321},
	2242
	(2006).
	
	
	
\bibitem{Raussendorf07}
	R. Raussendorf and J. Harrington,
	Phys. Rev. A
	\textbf{98},
	190504
	(2007).	
		
	

\bibitem{Raussendorf05}
	R. Raussendorf, S. Bravyi and J. Harrington,
	Phys. Rev. A
	\textbf{71},
	062311
	(2005).	



	
\bibitem{Jennings09}
	D. Jennings, A. Dragan, S. D. Barrett, S. D. Bartlett and T. Rudolph,
	Phys. Rev. A
	\textbf{80},
	032328
	(2009).
	



\bibitem{Aolita13}
	L. Aolita, F. G. S. L. Brand\~{a}o and A. J. Roncaglia,
	in preparation,
	(2014).
	
\bibitem{orieux}	
A. Orieux, J. Boutari, M. Barbieri, M. Paternostro and P. Mataloni,
arXiv:1312.1102
(2013).

\bibitem{kay06}
	A. Kay, J. K. Pachos, W. D\"ur and J. J. Briegel,
	N. Jour. Phys.
	\textbf{8},
	147
	(2006).
	
	
\bibitem{dcavalcanti10}
	D. Cavalcanti, L. Aolita, A. Ferraro and A. Ac\'\i{}n,
	New J. Phys.
	\textbf{12},
	025011
	(2010).
	
	
	
	
	
\bibitem{walther12}
	A. Aspuru-Guzik and P. Walther
	Nat. Phys.
	\textbf{8},
	285
	(2012).
	
	
\bibitem{kay10}
	A. Kay,
	J. Phys. A: Math. Theor.
	\textbf{43},
	495301
	(2010).
	
	
\bibitem{kwiat99}
	P. G. Kwiat, E. Waks, A. G. White, I. Appelbaum and P. H. Eberhard,
	Phys. Rev. A
	\textbf{60},
	R773
	(1999).
		
\bibitem{james01}
	D. F. V. James, P. G. Kwiat, W. J. Munro and A. G. White,
	Phys. Rev. A
	\textbf{64},
	052312
	(2001).
	
	
\bibitem{farias12b}
	O. Jim\'enez Far\'ias, G. H. Aguilar, A, Vald\'es-Hern\'andez, P. H. Souto Ribeiro, L. Davidovich, and Walborn, S. P.
	Phys. Rev. Lett.
	\textbf{109},
	150403
	(2012).	
	
	
	
\bibitem{vidal02}
	G. Vidal and R. F. Werner,
	Phys. Rev. A
	\textbf{65},
	032314
	(2002).
	
	
\bibitem{dcavalcanti09}
	D. Cavalcanti, R. Chaves, L. Aolita, L. Davidovich,  and A. Ac\'\i{}n,
	Phys. Rev. Lett..
	\textbf{103},
	030502
	(2009).	


\bibitem{Aolita_graph_2010}
	L. Aolita, D. Cavalcanti, R. Chaves, C. Dhara,  L. Davidovich,  and A. Ac\'\i{}n,
	Phys. Rev. A
	\textbf{82},
	032317
	(2010).
	
	
	
	
	
\bibitem{Chaves_noisy_2011}
	R. Chaves and F. de Melo,
	Phys. Rev. A
	\textbf{84},
	022324
	(2011).
	
	
	
\bibitem{Renes04}
	J. M. Renes and R. Blume-Kohout and A. J. Scott and C. M. Caves,
	J. Math. Phys.
	\textbf{45},
	2171
	(2004).
	
	
	


	
\bibitem{Klappenecker05}
	A. Klappenecker and M. Roetteler
	Proceedings 2005 IEEE International Symposium on Information Theory (ISIT 2005), Adelaide, Australia,
	1740
	(2005).
	
	
	
\bibitem{Horodecki1999}
	M. Horodecki, P. Horodecki  and  R. Horodecki,
	Phys. Rev. A
	\textbf{60},
	1888
	(1999).
		
\bibitem{amselem09}
	E. Amselem and M. Bourennane,
	Nat. Phys.
	\textbf{5},
	748
	(2009).
		
		
\bibitem{lavoie10}
	J. Lavoie, R. Kaltenbaek, M. Piani, and K. J. Resch,
	Phys. Rev. Lett.
	\textbf{105},
	130501
	(2010).
	
	
	
\bibitem{barreiro10}
	J. T. Barreiro,  P. Schindler, O. G\"uhne, T. Monz,  M. Chwalla, C. F. Roos, M. Hennrich and R. Blatt,
	Nat. Phys.
	\textbf{6},
	943
	(2010).
	
		
\bibitem{DiGuglielmo2011}
	J. DiGuglielmo,  A. Samblowski, B. Hage, C. Pineda, J. Eisert, and R. Schnabel,
	Phys. Rev. Lett.
	\textbf{107},
	240503
	(2011).
	
	\bibitem{Kaneda2012}
	F. Kaneda, R. Shimizu, S. Ishizaka, Y.  Mitsumori, H. Kosaka, and K. Edamatsu,
	Phys. Rev. Lett.
	\textbf{109},
	040501
	(2012).
	
		
	
\bibitem{Hiesmayr2013}
	B. C. Hiesmayr and W. L\"{o}ffler,
	New J. Phys.
	\textbf{15},
	083036
	(2013).
	

	
	
						
\end{thebibliography}


\end{document}